\documentclass[a4paper,fleqn,usenatbib]{mnras}

\usepackage{newtxtext,newtxmath}

\usepackage[T1]{fontenc}
\usepackage{ae,aecompl}

\usepackage{graphicx}	
\usepackage{amsmath}	
\usepackage{amssymb}	

\def\M1{M_1^\prime}

\title[Differing Relations of Centrals \& Satellites]{The Differing Relationships Between Size, Mass, Metallicity and Core Velocity Dispersion of Central and Satellite Galaxies}

\author[A. Spindler et al.]{
Ashley Spindler,$^{1}$\thanks{E-mail: Ashley.Spindler@open.ac.uk (AS)}
David Wake,$^{1,2}$
\\
$^{1}$Department of Physical Sciences, The Open University, Milton Keynes, MK7 6AA, UK\\
$^{2}$Department of Astronomy, University of Wisconsin-Madison, Madison, WI 53706, United States
}

\date{Accepted XXX. Received YYY; in original form ZZZ}

\pubyear{2016}

\begin{document}
\label{firstpage}
\pagerange{\pageref{firstpage}--\pageref{lastpage}}
\maketitle

\begin{abstract}
We study the role of environment in the evolution of central and satellite galaxies with the Sloan Digital Sky Survey. We begin by studying the size-mass relation, replicating previous studies, which showed no difference between the sizes of centrals and satellites at fixed stellar mass, before turning our attention to the size-core velocity dispersion ($\sigma_0$) and mass-$\sigma_0$ relations. By comparing the median size and mass of the galaxies at fixed velocity dispersion we find that the central galaxies are consistently larger and more massive than their satellite counterparts in the quiescent population. In the star forming population we find there is no difference in size and only a small difference in mass. To analyse why these difference may be present we investigate the radial mass profiles and stellar metallicity of the galaxies. We find that in the cores of the galaxies there is no difference in mass surface density between centrals and satellites, but there is a large difference at larger radii. We also find almost no difference between the stellar metallicity of centrals and satellites when they are separated into star forming and quiescent groups. Under the assumption that $\sigma_0$ is invariant to environmental processes, our results imply that central galaxies are likely being increased in mass and size by processes such as minor mergers, particularly at high $\sigma_0$, while satellites are being slightly reduced in mass and size by tidal stripping and harassment, particularly at low $\sigma_0$, all of which predominantly affect the outer regions of the galaxies.
\end{abstract}




\section{Introduction}

The effect of environment on galaxy evolution has been an area of much study in recent times. Large surveys such as the Sloan Digital Sky Survey \citep[SDSS,][]{2000AJ....120.1579Y,2009ApJS..182..543A} provide astronomers with sufficient data to study galaxies statistically in a variety of environments. It is largely accepted that galaxies are situated in dark matter haloes which grow hierarchically over time, such that smaller haloes infall towards and merge with larger ones. These dark matter haloes play host to individual galaxies, but also to groups and clusters made up of pairs, tens or even thousands of galaxies \citep[]{2012ARA&A..50..353K, 1997ApJ...490..493N}.

Galaxies in different haloes and at different positions within those haloes have varying properties, which are related to the properties of, and location within, their host haloes. For example satellite galaxies experience lower levels of star formation than central galaxies due to quenching mechanisms, which predominantly act on those satellites \citep[]{2004MNRAS.353..713K, 2009MNRAS.394.1213W, 2013MNRAS.432..336W, 2006MNRAS.373..469B}. Indeed, two very similar galaxies could evolve in significantly different ways if placed in different environments; for example if one were to become a satellite galaxy while the other remains a central or isolated galaxy.

Scaling relations can tell us much about the evolution of galaxies and help constrain models of that evolution. The size-mass relation of early-type galaxies has been shown to be invariant with environment by a number of authors. \cite{2013ApJ...779...29H} showed that the size-mass relation of early type galaxies is the same for central and satellite galaxies, implying no environmental dependance. \cite{2014MNRAS.439.3189S} found no environmental dependance of bulge-dominated galaxy sizes at fixed mass in contrast to state-of-the-art semi-analytic models of galaxy formation which predicted a $\sim$1.5-3 times increase in size from low to high mass haloes.  \cite{2014MNRAS.439.3189S} describes a number of mechanisms within the models that predict an environmental dependance, such as violent disk instabilities which can lead to bulge growth \cite[]{2009ApJ...703..785D, 2011ApJ...730....4B, 2011ApJ...741L..33B}, minor mergers which happen up to $\sim$4 times more often in high mass haloes in some models \cite[]{2012MNRAS.424..361M, 2013MNRAS.433.1479H}, gas dissipation following mergers which would more effectively shrink galaxies in less massive haloes and finally satellite evolution which could stunt the growth of central galaxies from mergers if satellite star formation is quenched in a fast manner as opposed to a slow one.

In addition to the size-mass relation \cite{2012MNRAS.425..296M} showed that the Tully-Fisher relation for disk galaxies is largely unaffected by environment, finding only a small steepening with increasing local density. \cite{2015RAA....15..651H} find a significant environmental dependance on the fundamental plane coefficients of early-type central galaxies in SDSS DR7 (Data Release 7), but find no dependance on environment for the same coefficients of satellite galaxies.

The preservation of the size-mass relation for centrals and satellites is an important test of environmental processes in galaxy formation models, however it does not completely rule out environmental processes modifying the size and mass for a given galaxy. Previous studies have compared centrals and satellites at fixed stellar mass in an attempt to control for the strong dependency of many galaxy properties on mass, trends which are largely independent of environment \cite[]{2008arXiv0805.0002V}. However, to investigate how the evolution of a galaxy differed as a result of it's host halo being accreted onto a more massive halo, we must compare central and satellite galaxies that were the same before the satellites were accreted. Many of the physical processes that may be applied to a galaxy once it has become a satellite, such as stripping of its hot halo gas, ram pressure stripping of its cold gas, tidal stripping of its stars, and a modified merger rate, would affect its stellar mass evolution \cite[]{1972ApJ...176....1G, 1980ApJ...237..692L, 1999MNRAS.308..947A, 2003MNRAS.341..326D, 2004ApJ...615L.101B, 2015AN....336..505P, 2008MNRAS.387...79V}. As such, comparing centrals and satellites at fixed stellar mass will be unlikely to lead to a direct comparison of galaxies that were similar before the satellites became satellites. In many cases the physical processes that we wish to understand are exactly those processes that may affect the stellar mass evolution.

These physical processes could be acting in a way which preserves the size-mass relation, in essence hiding their effects from studies looking at this relation. For example, central galaxies are much more likely to experience minor mergers than satellites; this could have a `puffing up' effect by adding mass predominantly to the outer regions of the galaxy making them larger and more massive at the same time \cite[]{2010ApJ...709.1018V, 2009ApJ...699L.178N}. Satellites on the other hand can either have their star formation quickly halted by quenching, leading to no further growth or have their mass stripped away by tidal interactions with their dark matter halo and harassment by nearby galaxies.

Similar problems exist when comparing galaxy populations at different epochs, which have also typically been studied at fixed stellar mass. Recently a number of groups have tried to more directly study progenitor and descendant populations at differing masses by using fixed cumulative stellar mass number density selections \cite[e.g][]{2007ApJ...655L..69W,2008MNRAS.387.1045W,2010ApJ...709.1018V,2011MNRAS.412.1123P,2013ApJ...778..115P,2014ApJ...780...34L}, matched galaxy population from halo modelling and subhalo abundance matching (SHAM) \cite[e.g.][]{2007ApJ...667..760Z,2007ApJ...655L..69W, 2008MNRAS.387.1045W,2008ApJ...679.1192C,2009ApJ...696..620C,2013ApJ...770...57B,2013MNRAS.428.3121M}, or by comparing populations at fixed core velocity dispersion \cite[]{2012ApJ...760...62B}. Of these techniques, number density selection cannot be applied to the central satellite comparison. Halo modelling and SHAM has been used to remove the effect of differing stellar mass growth when comparing centrals and satellites in the work of \cite{2013MNRAS.432..336W}. In this paper we will make comparisons at fixed core velocity dispersions ($/sigma_0$) for central and satellite galaxies. Core velocity dispersion is largely invariant to growth by minor mergers \cite[e.g.][]{2003ApJ...589...29L,2012ApJ...760...62B} and internal processes that could significantly change $/sigma_0$, such as major bursts of star formation or puffing up via mass loss from quasars  \cite[]{2008ApJ...689L.101F}, are rare at the redshifts of this study. Indeed \cite{2015MNRAS.454.2770T} show that in the Illustris simulation the core velocity dispersion of the most massive progenitors of z=0 galaxies has hardly changed on average since $z\sim1.5$.  Furthermore, since the cores of the galaxies are shielded from the environment by their outer regions, $\sigma_0$ should remain largely unaffected by the possible mechanisms which could drive or prevent growth such as minor mergers, star formation, or tidal stripping. We compare the sizes and masses of central and satellite galaxies selected from the Sloan Digital Sky Survey and study the radial mass distribution of the galaxies, relative to the core velocity dispersion.

In Section \ref{Data} we describe the data sample and the cuts made. In Section \ref{results} we revisit the size-mass relation for satellite and central galaxies, and investigate the size-$\sigma_0$ and mass-$\sigma_0$ relations. We also study the mass distributions of the central and satellite galaxies in bins of core velocity dispersion and discuss the implications of these results. Finally in Section \ref{Conclusions} we discuss and summarise our results. Throughout the paper we consider a standard cosmology with the following parameters: $H_0 = 100$ $h$ km s$^{-1}$, $\Omega_0 = 0.3$ and $\Omega_{\Lambda} = 0.7$.

\section{Data}
\label{Data}

The galaxies used in this analysis come from the main sample of the SDSS Data Release 7 \citep[DR7,][]{2009ApJS..182..543A}. We combine the Large Scale Structure Sample of the New York University Value Added Galaxy Catalog \citep[VAGC,][]{2005AJ....129.2562B}, the MPA/JHU catalog \cite[]{2003MNRAS.341...54K, 2004MNRAS.351.1151B, 2007ApJS..173..267S}, the UPenn SDSS PhotDec Catalog \citep[UPenn,][]{2015MNRAS.446.3943M} and the Yang Group Catalog \citep[]{2009ApJ...695..900Y}. The catalogs have a coverage of 7966 sq degrees, a r-band magnitude range of $-23.5 < M_r < -19.5$. We apply a redshift cut of $0.01< z < 0.1$, which leaves 291,042 galaxies in the sample.

From the VAGC we take the fibre measurements for the Velocity Dispersion ($\sigma$) and the redshifts (Z). The velocity dispersions from the VAGC are taken from inside the 3 arc second fibre of SDSS and need to be corrected to represent the core velocity dispersion of the galaxy. To do this we corrected $\sigma$ to be within one-eighth of the $r_e$ of the galaxy using the following relationship: $\sigma_0 = \sigma_{ap}(8r_{ap}/r_e)^{0.066}$ \citep[]{2006MNRAS.366.1126C}. As $\sigma_0$ has large errors at both high and low values, we only consider galaxies with $70 kms^{-1} < \sigma < 300 kms^{-1}$ and $\Delta\sigma < 15\%$.

From MPA/JHU catalog we use the specific star formation rates ($sSFR = SFR/M_*$), which are calculated using the technique described in \citet{2004MNRAS.351.1151B}. We also take the stellar masses and r and g-band magnitudes, which we use to fit a stellar mass-to-light ratio \cite[]{2003MNRAS.341...54K}.

The UPenn PhotoDec Catalog uses two component fits of the photometry to estimate a variety of galaxy properties, as opposed to the single component fits of earlier catalogs, we use the best fit for each galaxy which may be a single component or a combination of S\'ersic, exponential and De Vaucouleurs profiles. We use the total r and g-band magnitudes to calculate new stellar masses from the best fits of the catalog. In addition we use the best fit half-light radii; galaxies with $r_e > 30$ kpc are cut from the sample. We also use the axis ratios from this catalog to cut out edge-on spirals, as the rotational velocity of these galaxies can contaminate their core velocity dispersion and cause it to appear higher than it would otherwise be. To achieve we remove galaxies with a $b/a$ ratio $< 0.1$. Finally we use the S\'ersic indices from the best single component fits.

We calculate the maximum volume, $V_{max}$, over which a galaxy could be observed using the flux limit of SDSS, the mass of the galaxy, the redshift limit and the mass completeness limits as derived from \cite{2012ApJ...751L..44W}. This process corrects for the fact that many low brightness galaxies are not picked up by SDSS as they fall below the flux limit, which would skew the results at higher redshifts to high luminosity galaxies.

Applying the cuts laid out in this section gives a final sample of 124,524 galaxies. The majority of the galaxies not included in our final sample are disqualified by the stellar mass completeness limit and velocity dispersion cuts.

\section{Defining Central and Satellite Galaxies}
\label{censat}

To define central and satellite galaxies we primarily use the Yang Group Catalog \citep[]{2009ApJ...695..900Y}, which uses an iterative group finding algorithm to construct dark matter haloes from the SDSS galaxies. Starting with a simple friends-of-friends algorithm, tentative groups are found and assigned properties based on their characteristic luminosities. Using the properties of these tentative groups the algorithm decides whether or not to add galaxies from the nearby redshift space. If new galaxies are added to the group, the dark matter properties are recalculated based on the new group luminosity. The group finder iterates over this process until no changes to the groups are made. 

The Yang Catalog has two definitions for the centrals and satellites, these are based on the stellar mass and luminosity of the galaxies in each halo. In these definitions the most massive or brightest galaxies are designated the central and the rest of the galaxies satellites. We choose to use the mass rather than luminosity definition for our centrals, but take some further steps to mitigate any bias that this definition may introduce into our analysis. As we have selected centrals to be more massive, this could introduce a small bias into the mass-$\sigma_0$ relation (and by extension the size-$\sigma_0$ relation), making the central galaxies more massive at fixed $\sigma_0$. Consider a halo where the two most massive galaxies have identical central velocity dispersions; the most massive galaxy will be designated the central and the least massive the satellite. As a result central galaxies may appear more massive than satellite galaxies at fixed $\sigma_0$ simply as a result of the central/satellite definition. This definition could also introduce other similar biases for any galaxy property that depends on stellar mass.

To mitigate this potential bias we consider throughout the paper two additional central/satellite definitions. The first chooses to use the galaxy with the highest $\sigma_0$ as the central in all haloes. This leads to approximately 2500 centrals (~5\% of the groups in the sample) being swapped with a satellite from their halo that has a higher dispersion. The second combines the mass and $\sigma_0$ definitions by randomly selecting which galaxy (the highest mass or the highest dispersion) will be the central, resulting in roughly 1250 centrals being swapped with satellites. We use this random sampling method to define the main sample throughout this work, but we also show the results for both the mass and dispersion samples for reference. We note that whilst the three definitions of central galaxies produce slightly different results in the manner that we expected, using any of them would result in the same conclusions for all the trends that we study. The same is true for the most conservative method of only using haloes where the central galaxy has both the highest stellar mass and $\sigma_0$, although we do not make use of this definition since it removes a large number of galaxies from our sample.
\begin{figure*}
\includegraphics[trim = 40mm 20mm 30mm 20mm, clip, width=1.0\textwidth]{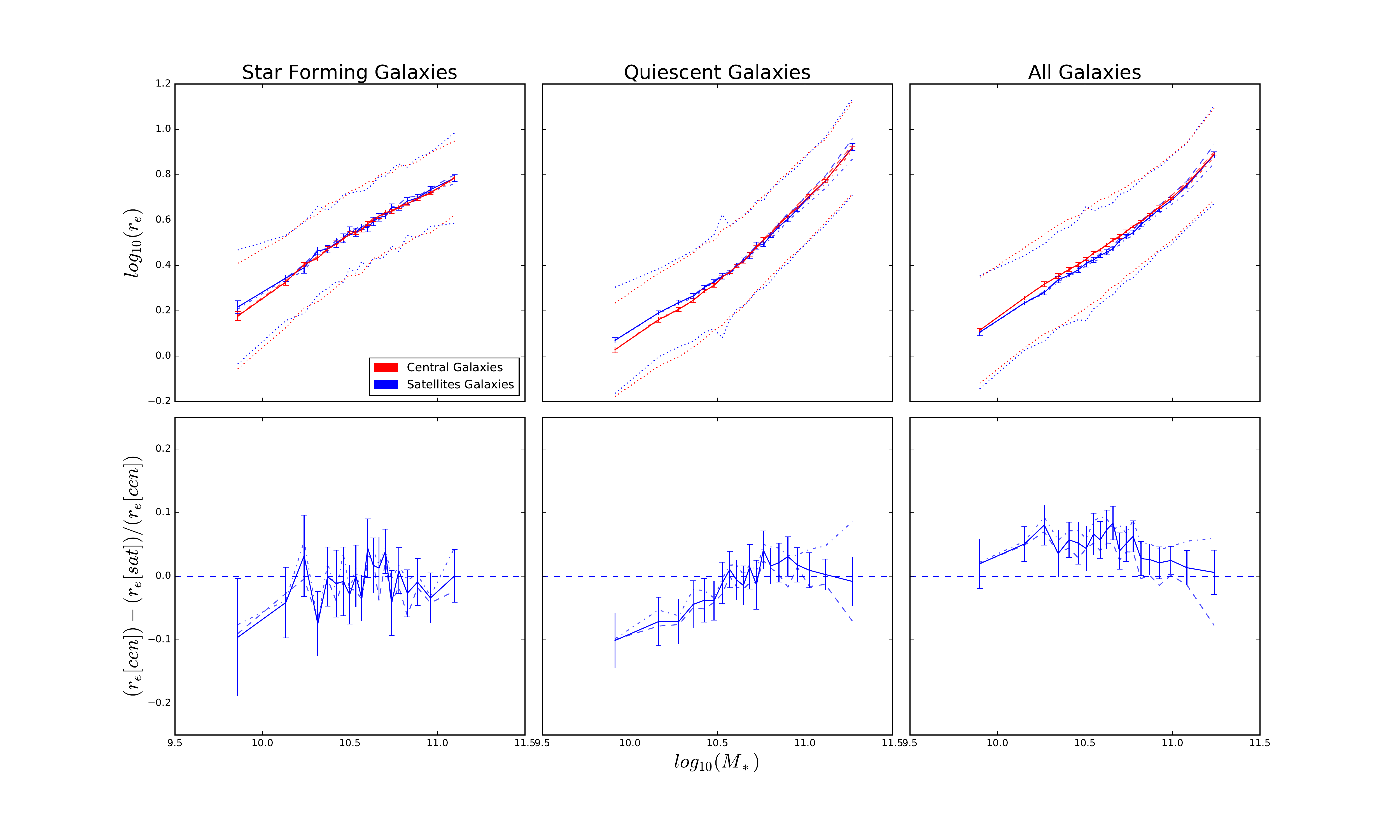}
\centering
\caption{\small
(Top) Comparison of the median half-light radius at fixed mass of central (red line) and satellite (blue line) galaxies. Left is the star forming population, centre is the quiescent and right is the entire population. The dotted lines represent the 1$\sigma$ scatter. The solid lines represent the random central/satellite split, the faint dot-dashed lines represents the mass split and the faint dashed lines represent the $\sigma_0$ split (see the main text for details). (Bottom) The fractional difference in median radius at fixed mass for central and satellite galaxies. We see that for the star forming and quiescent population the size-mass relation is almost the same for centrals and satellites. For all galaxies the centrals are slightly larger than the satellites, but this is predominantly due to the different quiescent fractions in the central and satellite populations (see text).
\label{fig:MassRadius}}
\end{figure*}

\begin{figure*}
\includegraphics[trim = 40mm 20mm 30mm 20mm, clip, width=1.0\textwidth]{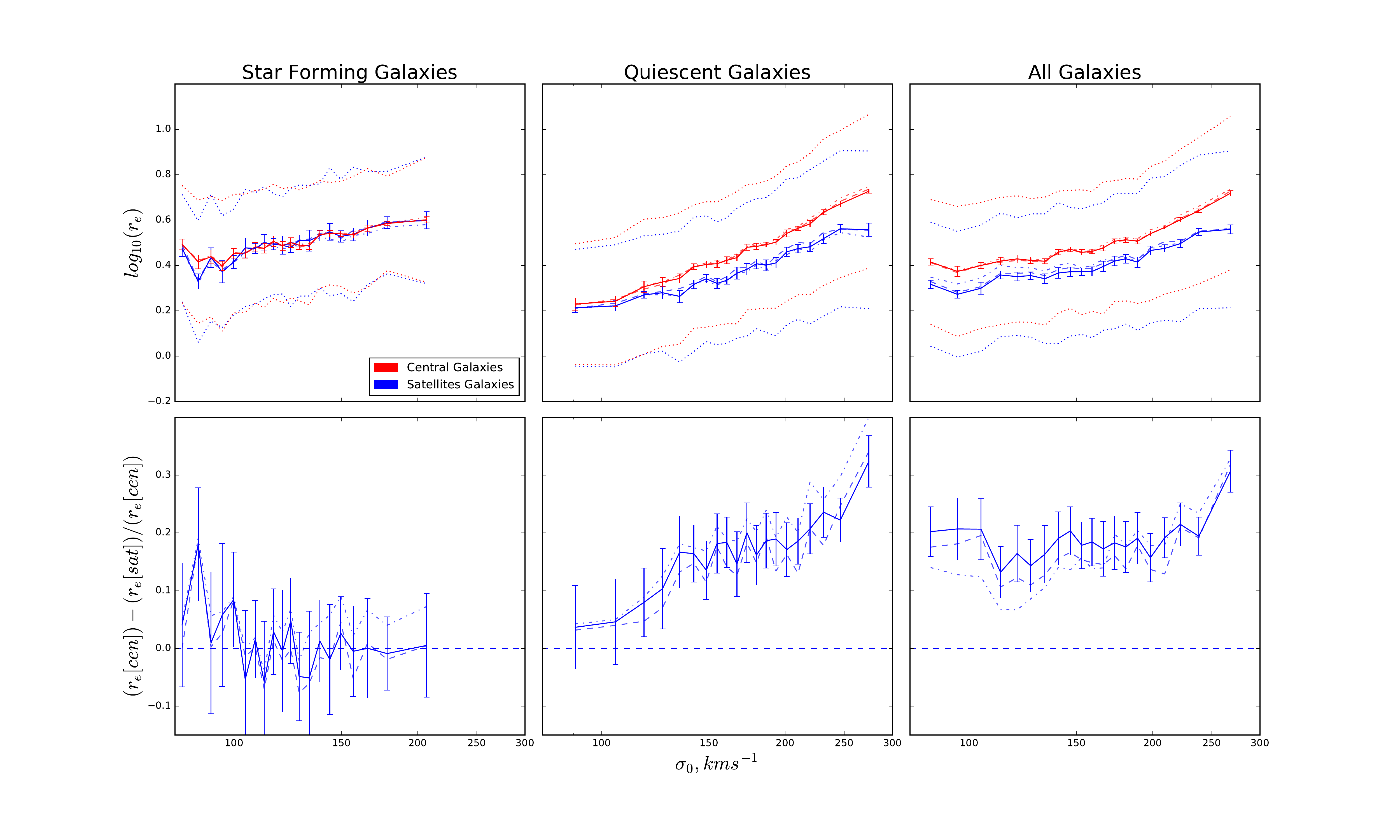}
\centering
\caption{\small
(Top) Comparison of the median half-light radii of central (red) and satellite (blue) galaxies. Left is the star forming population, centre is the quiescent population and right is the entire population. The dashed lines represent the 1$\sigma$ scatter. The solid lines represent the random central/satellite split, the faint dot-dashed lines represents the mass split and the faint dashed lines represent the $\sigma_0$ split (see the main text for details). (Bottom) The fractional difference in median radius at fixed $\sigma_0$ for central and satellite galaxies.  For the star forming population we see no difference in size at fixed central dispersion, but for quiescent galaxies we see that the centrals become increasingly larger at higher dispersions than for satellites. For all galaxies there is a constant difference in size.
\label{fig:VdispRadius}}
\end{figure*}

\begin{figure*}
\includegraphics[trim = 40mm 20mm 30mm 20mm, clip, width=1.0\textwidth]{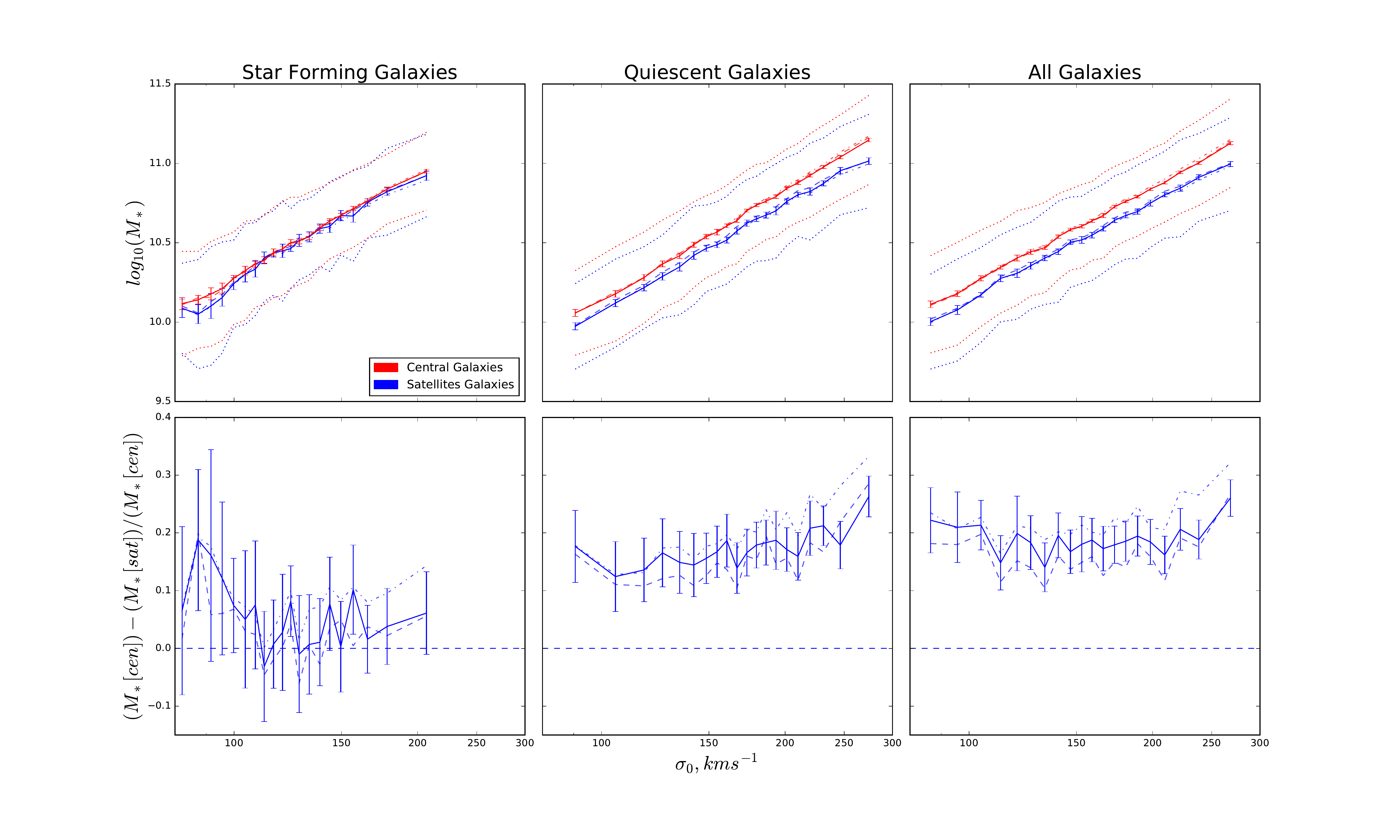}
\centering
\caption{\small
(Top) Comparison of the median stellar mass of central (red) and satellite (blue) galaxies. Left is the star forming population, centre is the quiescent group and right is the entire population. The dashed lines represent the 1$\sigma$ scatter. The solid lines represent the random central/satellite split, the faint dot-dashed lines represents the mass split and the faint dashed lines represent the $\sigma_0$ split (see the main text for details). (Bottom) The fractional difference in median mass at fixed $\sigma_0$ for central and satellite galaxies. The star forming population shows a small difference in mass at fixed dispersion and the quiescent population has a larger difference in mass, which increases with higher dispersion. For all galaxies the difference in mass remains roughly constant.
\label{fig:VdispMass}}
\end{figure*}
\section{Results}
\label{results}
\subsection{The Size-Mass Relation}

We begin by comparing the size-mass relation of the central and satellite galaxies. In Figure \ref{fig:MassRadius} we plot the median $r_e$ of the centrals and satellites at fixed mass in 20 stellar mass bins, each of which has the same number of satellite galaxies. The top row is the size-mass relation and the bottom row is the difference between the centrals and satellites. The errors are calculated from 1000 bootstrap resamplings. In all cases we weight by V$_{max}$. The panels from left to right  show relations for star forming galaxies, quiescent galaxies, and all galaxies. We split the galaxies into star forming and quiescent populations by applying a cut at $log_{10}(sSFR) = -11$, which sits at the minimum between the two peaks in the bimodal sSFR distribution \cite[e.g.][]{2012MNRAS.424..232W}. 

We apply this sSFR cut for two reasons. Firstly, as is visible in Figure \ref{fig:MassRadius}, star forming galaxies typically have larger $r_e$ than quiescent galaxies at the same mass (or $\sigma_0$) although, the difference decreases at higher masses and dispersions. This difference most likely results from extended luminous star forming disks in the star forming galaxies increasing the $r$-band $r_e$. If we ignored this dependence on star formation, mean differences in the sizes of centrals and satellite galaxies could result from differences in the quiescent fractions, which are known to be higher in satellite galaxies at all masses \cite[e.g.][]{2012MNRAS.424..232W}. Secondly, since a central galaxy is more likely to be star forming than a satellite, we would like to separate out the effects of ongoing star formation, which would preferentially produce more massive central galaxies, from processes like tidal stripping of stars and minor mergers, which act on SF or quiescent galaxies alike.

The solid lines in Figure \ref{fig:MassRadius} represent the random central/satellite sample, while the faint dot-dashed and dashed lines represent the samples with centrals defined by mass and dispersion respectively. The three different central definitions show very similar relationships with the largest divergence at the high mass end. This plot mimics those found in \cite{2013ApJ...779...29H} (and others), that the size-mass relation is largely the same for centrals and satellites. 

Even though there is very little difference between the central and satellite size-mass relations for either star forming or quiescent galaxies there is a small difference in size for the full galaxy population, with central galaxies being larger at all masses. This larger difference is a result of the differing quiescent fractions of central and satellite galaxies. The higher fraction of quiescent satellites at all masses results in smaller sizes relative to the centrals, which are more likely to be star forming. 

The fact that the size-mass relation does has very little dependence on environment is an interesting result in and of itself (e.g. see \cite{2014MNRAS.439.3189S} for comparisons to semi-analytic galaxy formation model predictions), but it does not tell us how the sizes and masses of individual galaxies may be being changed by their transition from central to satellite relative to those that remain centrals. It only tells us that if the masses and sizes are changing for centrals and satellites these changes must be linked. For example, if mass is lost then the galaxy must shrink in $r_e$, and if mass is gained then the galaxy must expand in $r_e$, such that the size-mass relation remains unchanged. Another way to put this is that the stellar mass density of the centrals and satellites remains constant and satellites are no more or less compact than their central counterparts.

Minor mergers and ongoing star formation provide mechanisms for growth that predominately apply to central galaxies and which could increase the size of galaxies in a way which preserves the size-mass relation. However, tidal stripping and harassment mainly affect satellites and could lead to reduced sizes and masses, again in a way which could preserve the size-mass relation. \cite{2013MNRAS.432..336W} estimate that as a result of the quenching of their star formation quiescent satellite galaxies are up to 10\% less massive than they would have been had they remained centrals. Tidal stripping is likely to have further reduced the mass of satellites galaxies by up to 20\%, depending on the halo mass \cite[][]{2016arXiv160903379B}. For these reasons comparing galaxies at fixed mass may not be the best way to approach studying environmental effects on galaxies.  Instead we want to be able to compare central and satellite galaxies that were the same or similar prior to a satellites accretion onto a new dark matter halo.

\begin{figure*}
\includegraphics[trim = 40mm 20mm 30mm 20mm, clip, width=1.0\textwidth]{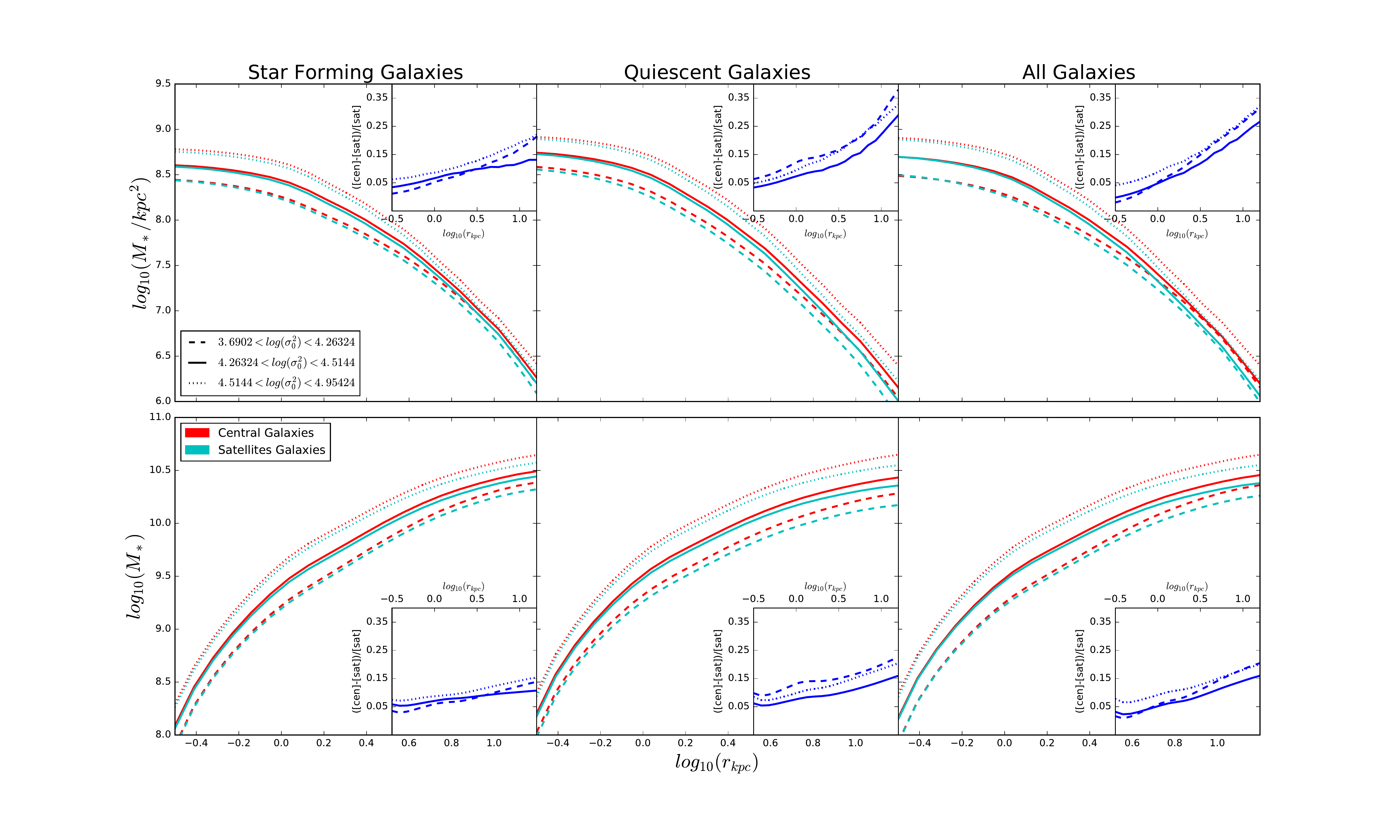}
\centering
\caption{\small
(Top) The mean surface mass density at fixed radius of the star forming, quiescent and full galaxy populations. (Bottom) The cumulative mean mass at fixed radii. The dashed, solid and dotted lines represent increasing bins of velocity dispersion, while the red lines are the central galaxies and the blue lines are the satellite galaxies. The inset figures show the fractional differences in mass density and cumulative mass at fixed radii. In the core regions of the galaxies we see that the centrals and satellites have the same mass density, indicating that the cores are largely unaffected by the environment. However in the outer regions of the galaxies the mass density begins to diverge, with centrals having more mass at larger radii. This relationship is stronger for quiescent galaxies than for star forming galaxies, which mirrors our previous results. We only consider the random central/satellite split in this Figure.
\label{fig:MagProfs}}
\end{figure*}

\subsection{Size and Mass comparisons at fixed $\sigma_0$}

To determine how the growth of satellites and centrals differ, we need to consider a variable which is not likely to be affected by the environmental processes experienced by a galaxies whether they are a central or satellite. Therefore, we choose to investigate the relationships of $M_*$ and $r_e$ with the core velocity dispersion, $\sigma_0$, which, as already discussed, we expect to remain largely unchanged by the physical processes acting on centrals and satellites.

Figure \ref{fig:VdispRadius} shows the $V_{max}$ weighted median half-light radius of central and satellite galaxies in the top row and the difference between the medians in the bottom row as a function of $\sigma_0$. We use the same method as in Figure \ref{fig:MassRadius} to split the sample into bins of $\sigma_0$. As with Figure \ref{fig:MassRadius} the solid lines represent the random central/satellite split and the faint dot-dashed and dashed lines represent the mass and dispersion based samples respectively.

For the star forming galaxy sample we see that there is no difference in the size-$\sigma_0$ relation for the random central/satellite split. The centrals from the mass definition are slightly larger than the satellites, but this is likely due to the bias mentioned in Section \ref{Data}. When we use the dispersion based sample, the satellites become slightly larger than the centrals at dispersions above $\sigma_0^2 = 125 kms^{-1}$. However, we could in fact be introducing the opposite bias here as we have deliberately decided against classifying the more massive, larger galaxies as centrals in some haloes.

Unlike the star forming sample, the quiescent galaxies show a large difference in size between the centrals and satellites in all three central/satellite splits. For the randomly defined sample the fractional difference increases with $\sigma_0$ from $4 \pm 7 \%$ at $\sigma_0 = ~90 km/s$, to $32 \pm 4 \%$ at $\sigma_0 = 270 km/s$ with an average difference of $16\%$. In the mass defined sample we see that the $\sigma_0$ dependance is stronger, with the difference in size reaching $40 \pm 4 \%$, with an average difference of $19\%$. The dispersion defined sample shows a lower maximum size difference of only $25\%$ and an average difference of $13\%$.

In the full galaxy population the difference in size remains relatively constant at all values of $\sigma_0$ with an average difference of $16\%$. The relation is fairly constant for the random and dispersion defined samples, but has a small $\sigma_0$ dependance in the mass defined sample.

Given that the central and satellite size-mass relations are the same and the size-$\sigma_0$ relations are not, we would expect to find a difference in mass at fixed dispersion between centrals and satellites. In Figure \ref{fig:VdispMass} we plot the median stellar mass in bins of velocity dispersion. As with the size-$\sigma_0$ relation we see that there is a mass deficit in the satellite galaxies. The star forming galaxies exhibit an average difference of $5\%$ in mass for the random central definition sample, the difference in slightly higher at $8\%$ for the mass defined sample and disappears completely for the dispersion defined sample.

Quiescent central galaxies show a consistently higher mass at fixed dispersion than quiescent satellite galaxies. The relation is relatively flat for the randomly defined sample, with an average difference of $17\%$. The average difference is lower for the dispersion defined sample at $14\%$. For the mass defined sample there is a stronger dependance on $\sigma_0$ with the fractional difference reaching $32\%$ in the highest dispersion bin, and with an average difference of $20\%$.

\begin{figure*}
\includegraphics[trim = 40mm 20mm 30mm 20mm, clip, width=1.0\textwidth]{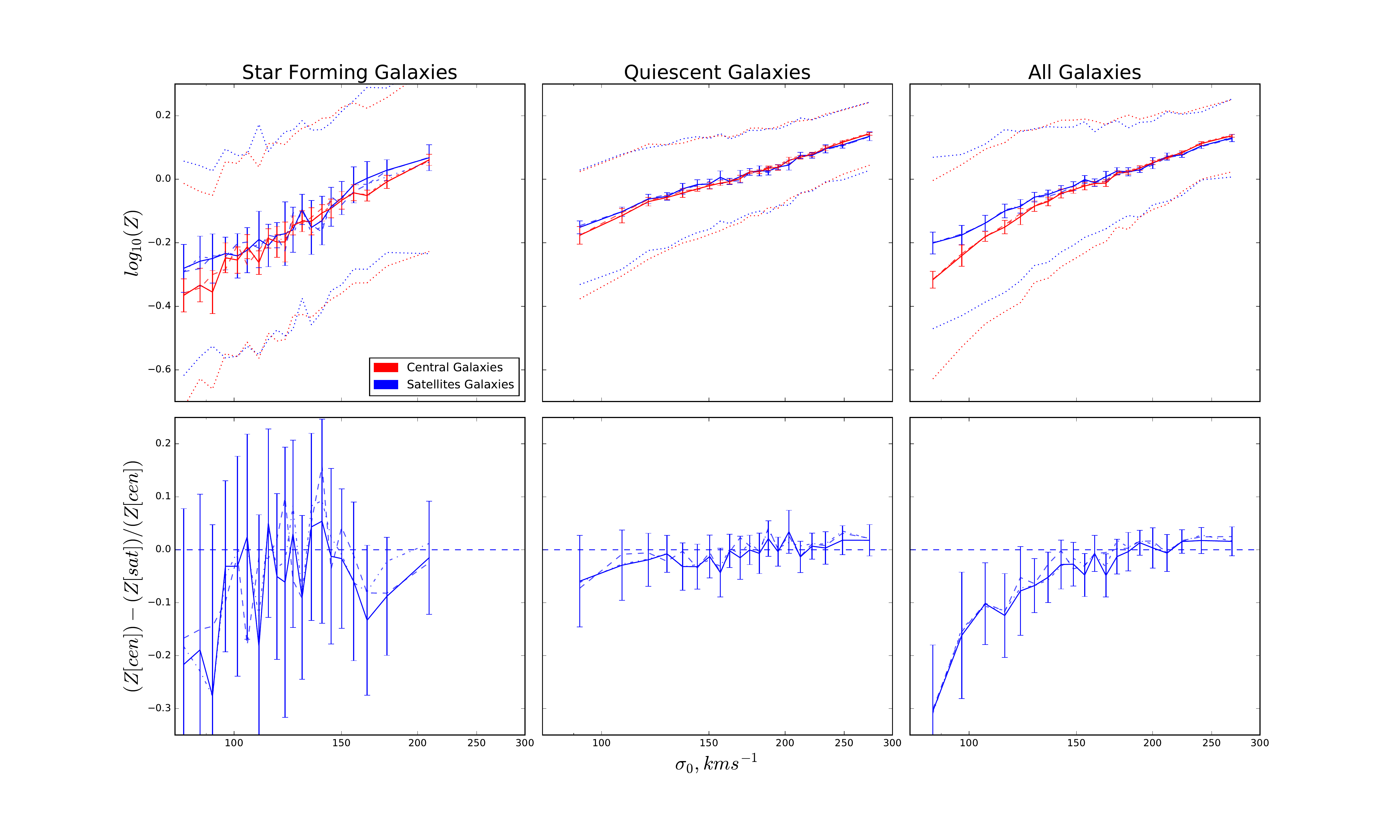}
\centering
\caption{\small
(Top) Comparison of the median stellar metallicity at fixed $\sigma_0$ of central (red line) and satellite (blue line) galaxies. Left is the star forming population, centre is the quiescent and right is the entire population. The dashed lines represent the 1$\sigma$ scatter. The solid lines represent the random central/satellite split, the faint dot-dashed lines represents the mass split and the faint dashed lines represent the $\sigma_0$ split (see the main text for details). (Bottom) The fractional difference in median stellar metallicity at fixed $\sigma_0$ for central and satellite galaxies. We see that for the star forming and quiescent population the metallicities for centrals and satellites are largely the same, except at very low dispersions for the star forming sample. Contrary to this, the all galaxy sample shows a strong trend with $\sigma_0$ in the difference in metallicity, where low dispersion centrals have lower metallicity than low dispersion satellites, this is however due to the relative mixing of star forming and quiescent centrals and satellites at different dispersions.
\label{fig:VoZ}}
\end{figure*}

\begin{figure*}
\includegraphics[trim = 40mm 20mm 30mm 20mm, clip, width=1.0\textwidth]{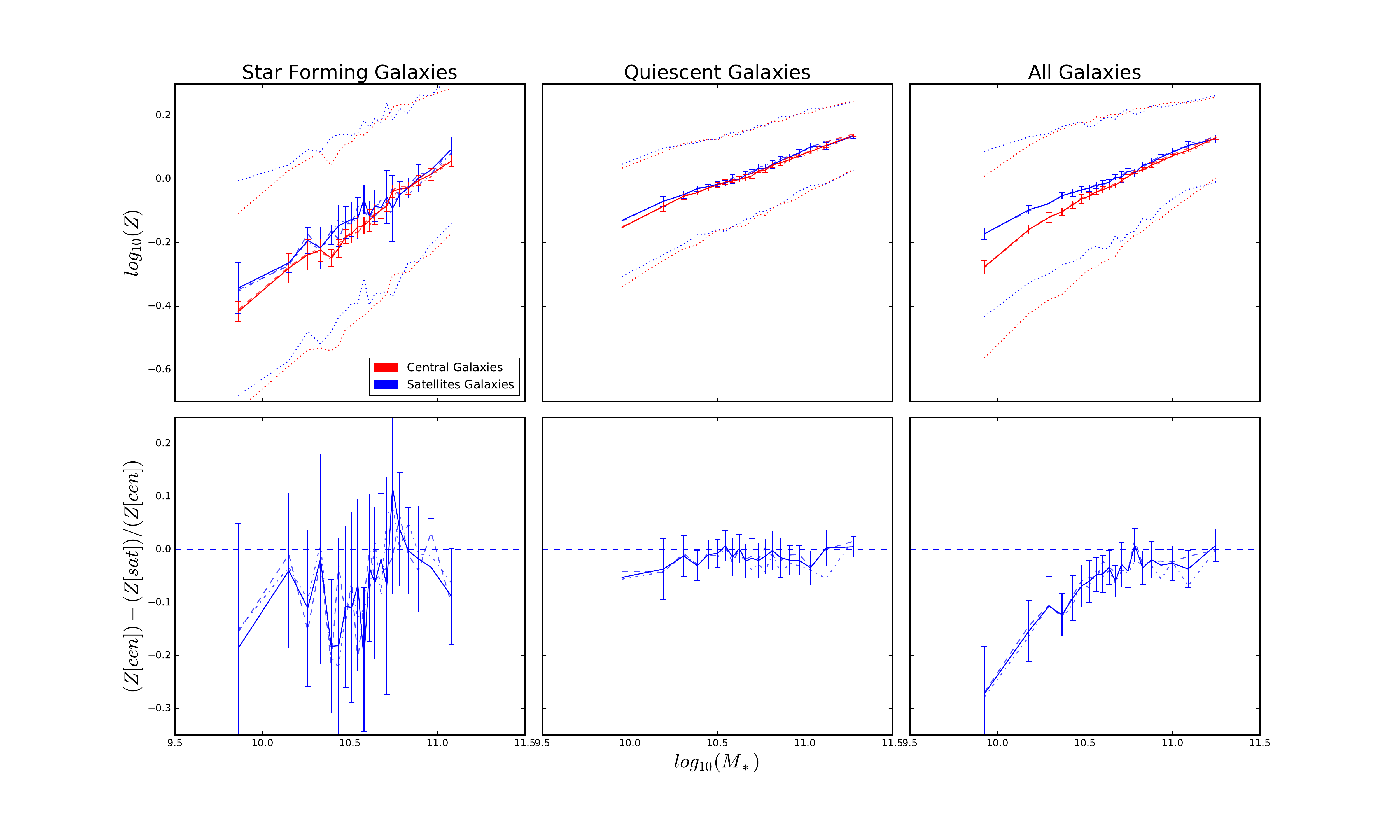}
\centering
\caption{\small
(Top) Comparison of the median stellar metallicity at fixed stellar mass of central (red line) and satellite (blue line) galaxies. Left is the star forming population, centre is the quiescent and right is the entire population. The dashed lines represent the 1$\sigma$ scatter. The solid lines represent the random central/satellite split, the faint dot-dashed lines represents the mass split and the faint dashed lines represent the $\sigma_0$ split (see the main text for details). (Bottom) The fractional difference in median stellar metallicity at fixed mass for central and satellite galaxies. As with the fixed $\sigma_0$ case we see little difference in the metal content of the centrals and satellites for the quiescent sample, and a small difference in the metal content of star forming centrals and satellites. As with the $\sigma_0$ there is a strong dependance on mass for the difference in metal content for the full galaxy population.
\label{fig:MassZ}}
\end{figure*}

\subsection{Radial Profiles}
\label{RadProfs}

If we assume that $\sigma_0$ is invariant with environment then the satellite and central galaxies were originally similar galaxies which have followed different evolutionary pathways. Our results indicate that this different evolution has either caused the centrals to continue growing, most likely through some combination of star formation and minor mergers, or that the satellites have been stunted by processes such as quenching, tidal stripping, and harassment. To investigate further we use the radial colour profiles in the $g$ and $r$ bands from SDSS to identify if there is a change in the mass distribution of the galaxies. If the environment is indeed playing a role in shaping the mass and size of the galaxies then we could see a difference in the mass profiles of the centrals and satellites. We might expect to see that the galaxy cores remain the same for centrals and satellites, indicating that the core and therefore the core velocity dispersion has not been affected by the environment. However, in the outskirts of the galaxies we might expect that the satellites will have less mass than the centrals, due to the satellites losing mass and the centrals gaining mass from the environment. We will investigate the mean radial profiles of surface mass density and total stellar mass.

We measure the radial mass profiles using the $r$-band and $g$-band profiles from SDSS. The magnitudes are k-corrected using the values from NYU-VAGC to a common redshift of $z=0.1$. To convert the magnitude profiles into mass profiles we require a mass-to-light ratio, which we estimate using the stellar masses of the galaxies and $g$- and $r$-band magnitudes from the UPenn catalog. We fit a linear relation to the observed $g-r$ colour and the stellar mass to $r$-band light ratios of all the galaxies in the sample. We fit the radial $g-r$ colour profiles of each galaxy to this relation to find the mass-to-light ratios in each of the radial bins. Finally, we use these mass-to-light ratios to convert the $r$-band magnitude profiles into stellar mass profiles.

In Figure \ref{fig:MagProfs} we plot the mean surface mass densities in units of $log_{10}(M_*/kpc^2)$ and cumulative mass profiles in units $log_{10}(M_*)$ for galaxies split in to three bins of velocity dispersion and into the three star formation rate populations. For a given star formation class surface mass density increases as the velocity dispersion increases. For a given dispersion the quiescent galaxies have higher surface mass densities than the star forming galaxies. We see the same trend in the cumulative mass profiles (bottom panels of Figure \ref{fig:MagProfs}), with the higher dispersion bin being the most massive and the quiescent galaxies being more massive than star forming galaxies. We also show the fractional differences in mass and mass density, which also clearly show the differences between centrals and satellites being more prominent at larger radii.

In each of the surface mass density plots we see that the centrals and satellites have very similar core densities. However, at larger radii the mass density of the centrals becomes higher than the satellites. Once again we see that the difference between the centrals and satellites is larger for the quiescent galaxies than for the star forming galaxies. The difference in density is also greater in the highest dispersion bin for quiescent galaxies, as was seen in the previous section. These results are reflected in the cumulative mass profiles. For a given bin in dispersion the satellite core masses are very close to the central core masses, but we see that most of the mass difference is in the outer regions of the galaxies.

\subsection{The Stellar Metallicity Relations}
\label{metals}

\cite{2010MNRAS.407..937P} found that satellite galaxies had lower stellar metallicity than centrals with the same stellar mass and that this metallicity difference increased as the stellar mass decreased. They argued that this could be being caused by mass stripping from the satellites, so that the satellites had the higher stellar metallicity of the higher mass galaxies they were before being stripped\footnote{\cite{2012MNRAS.425..273P} argue that this is not in fact the mechanism responsible based on gas phase metallicity measurement.}. The same result would also occur if the centrals were increasing in mass without changing their metalicity (e.g. from minor mergers), while the satellites remained unchanged. We revisit this relation and include results for the metallicity at fixed $\sigma_0$ and for galaxies split by star formation rate as before.

In Figures \ref{fig:VoZ} and \ref{fig:MassZ} we show the difference between central and satellite metallicity for SF, quiescent and all galaxies as a function of $\sigma_0$ and mass, respectively, in the same way as we have previously. We have used the stellar metallicity measurements of \cite{2005MNRAS.362...41G} as were used by \cite{2010MNRAS.407..937P}. One important aspect of these measurements is that the size of the SDSS fibre insures that the metallicity is being measured in the central part of the galaxy (typically within one $r_e$) and therefore should be largely unaffected by environmental processes in the same way $\sigma_0$ is unaffected, in particular minor mergers depositing metal poor stars on the outskirts of central galaxies.

One can see in Figure \ref{fig:VoZ} that for both SF and quiescent galaxies there is almost no difference between the metallicty of central and satellite galaxies at fixed $\sigma_0$. At fixed mass there is a small difference ($\sim$5\%) with central galaxies having lower metallicities than the equivalent satellites. In both cases there is a very weak trend such that the higher the mass or $\sigma_0$ of the centrals the smaller their deficit in metallicity relative to the satellites, with the highest $\sigma_0$ centrals actually being more metal rich. The $\sim$5\% difference in metallicity between centrals and satellites at fixed mass is consistent with the difference in mass measured at fixed $\sigma_0$ (Figure \ref{fig:VdispMass}) and the central mass-metallicty relation. In other words it is consistent with the central metallicity remaining unchanged between galaxies that become satellites or remain centrals and either central galaxies gaining mass and/or the satellite galaxies losing mass. The absence of any real difference in metallicity at fixed $\sigma_0$ further supports our tenet that $\sigma_0$ is unaffected by environmental processes.

It is worth pointing out a couple of other interesting aspects of Figure  \ref{fig:VoZ}. Since when SF and quiescent galaxies are considered separately, central and satellite galaxies show very little difference in metallicity at fixed mass or $\sigma_0$, the larger difference for the full galaxy population shown in the right hand panel of Figures \ref{fig:VoZ} and \ref{fig:MassZ}, and previously found by \cite{2010MNRAS.407..937P}, is largely being driven by differing fractions of SF and quiescent galaxies within the central and satellite population. There is a larger fraction of quiescent galaxies in the satellite population, which typically have higher metallicities than SF galaxies of the same mass or $\sigma_0$, causing the median metallicity of the satellites within the whole galaxy population to be higher. The increase in the difference between the central and satellite metallicity as mass or $\sigma_0$ decreases is caused by the combination of two effects; both the difference in metallicity between SF and quiescent galaxies and the difference in the fraction of quiescent galaxies between centrals and satellites increase as mass or $\sigma_0$ decrease. 

It's also worth noting that the lack of much of a difference in the stellar metallicity of central and satellite galaxies when split in SF and quiescent is at odds with the recent measurements from the EAGLE simulation \cite[]{2016arXiv160903379B}, which show a much larger difference in all star forming groups.

\section{Systematics}
\label{Systematics}

\begin{figure*}
\includegraphics[trim= 0mm 04mm 0mm 03mm, clip, width=1.0\textwidth]{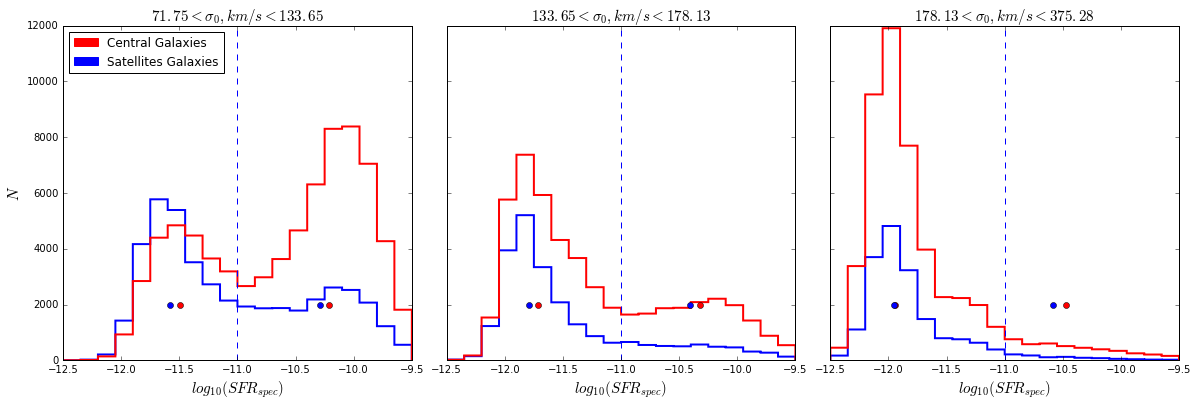}
\centering
\caption{\small
We show the distributions of Specific Star Formation Rate in three bins of core velocity dispersion. The Figure shows central galaxies in red and satellite galaxies in blue and the $log_{10}(sSFR) = -11$ cut for quiescent and star forming galaxies as the blue dashed line. We also mark the median sSFR for the centrals and satellites in both star formation groups with solid circles. We see that the distributions are similar, but the centrals have a higher median sSFR in most of the bins.
\label{fig:SFRVDisp}}
\end{figure*}

\begin{figure}
\includegraphics[trim= 0mm 22mm 0mm 20mm, clip, width=0.95\columnwidth]{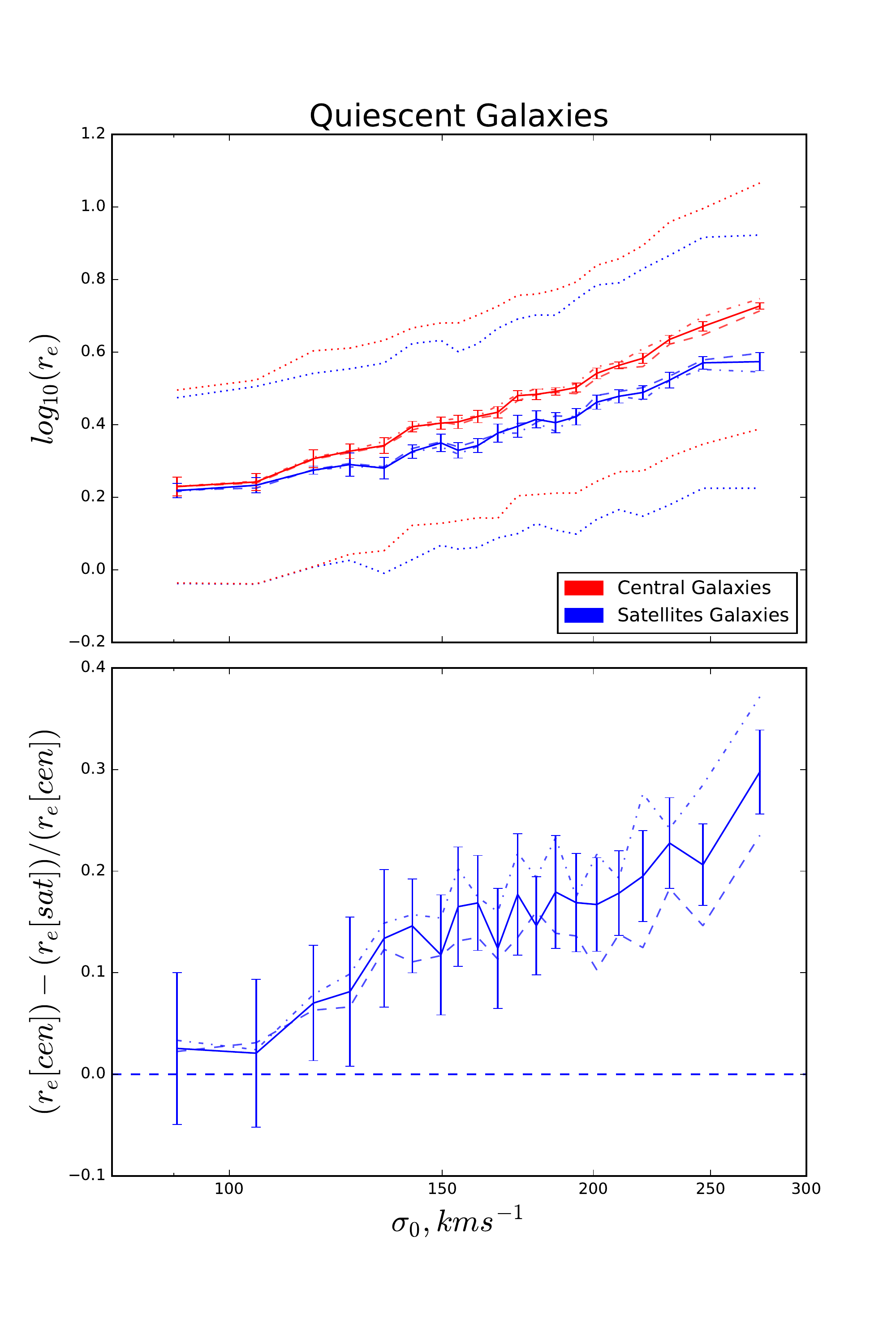}
\centering
\caption{\small
(Top) Comparison of the median half-light radii of central (red) and satellite (blue) galaxies in the quiescent group weighted by their relative specific star formation rate distributions (see the main text for details). The dashed lines represent the 1$\sigma$ scatter. The line styles are the same as in previous figures. (Bottom) The fractional difference in median radius at fixed $\sigma_0$ for central and satellite galaxies. We see that there is little difference between these weighted results and the results from Figure \ref{fig:VdispRadius}, indicating that differences between the central and satellite sSFR distributions are not responsible for the differences between central and satellite size and mass.
\label{fig:VoRadSFR}}
\end{figure}

\begin{figure*}
\includegraphics[trim= 0mm 03mm 0mm 03mm, clip, width=1.0\textwidth]{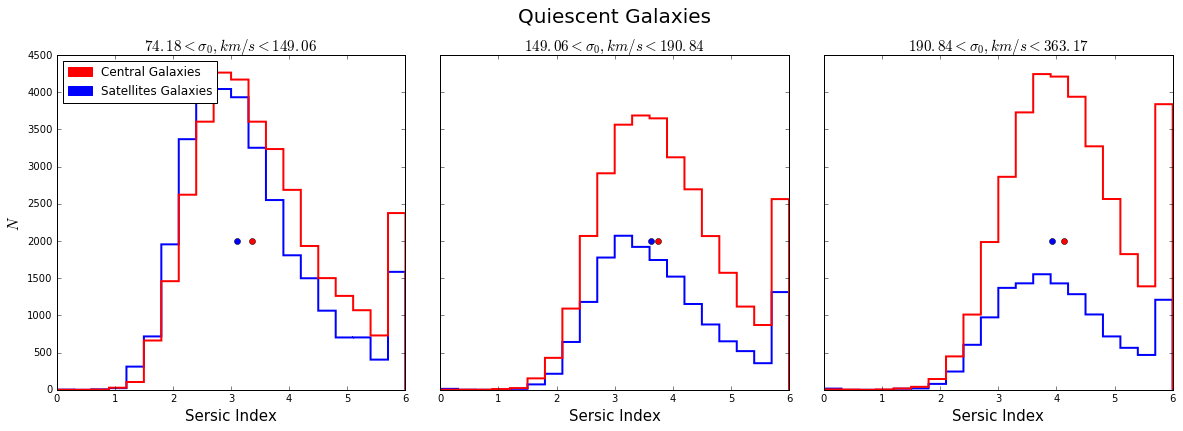}
\centering
\caption{\small
We show the distributions of S\'ersic Index in three bins of core velocity dispersion. The central galaxies are shown in red and satellite galaxies in blue. We also mark the median S\'ersic Index for the centrals and satellites in both star formation groups with solid circles. We see that the distributions are similar, but the centrals have a higher median S\'ersic Index in all of the bins. Note that there are a large number of galaxies in the S\'ersic Index = 6 bin as this is the maximum index assigned by the UPenn catalog.
\label{fig:SerVDisp}}
\end{figure*}

\begin{figure}
\includegraphics[trim= 0mm 22mm 0mm 20mm, clip, width=0.95\columnwidth]{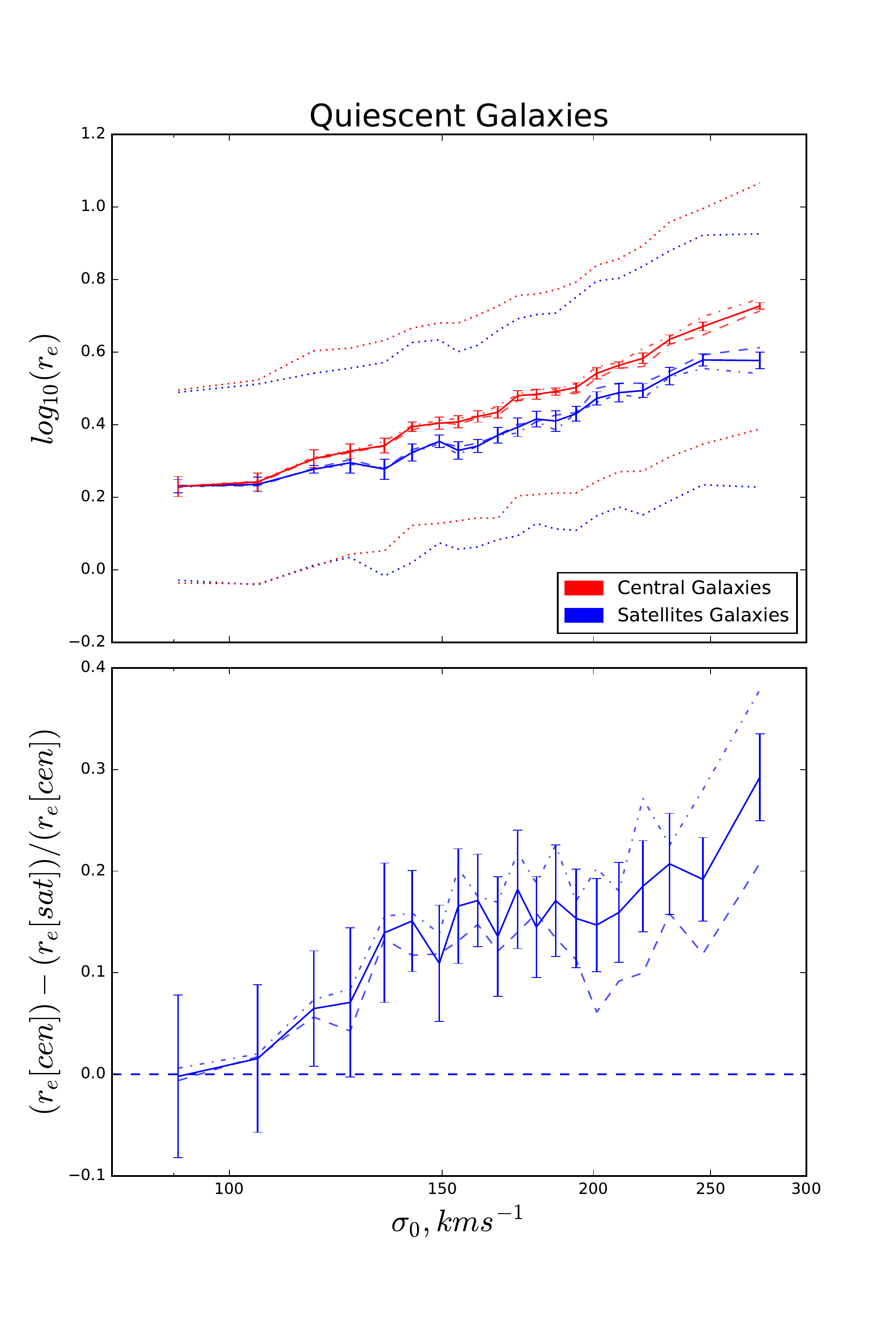}
\centering
\caption{\small
(Top) Comparison of the median half-light radii of central (red) and satellite (blue) galaxies in the quiescent group weighted by their relative S\'ersic Index distributions (see the main text for details). The dashed lines represent the 1$\sigma$ scatter. The line styles are the same as in previous Figures. (Bottom) The fractional difference in median radius at fixed $\sigma_0$ for central and satellite galaxies.  We see that the difference in size has decreased only slightly from the results in Figure \ref{fig:VdispRadius}, indicating that differences between the central and satellite morphology distributions are not responsible for the differences between central and satellite size and mass.
\label{fig:VoRadSer}}
\end{figure}

\begin{figure*}
\includegraphics[trim= 0mm 03mm 0mm 03mm, clip, width=1.0\textwidth]{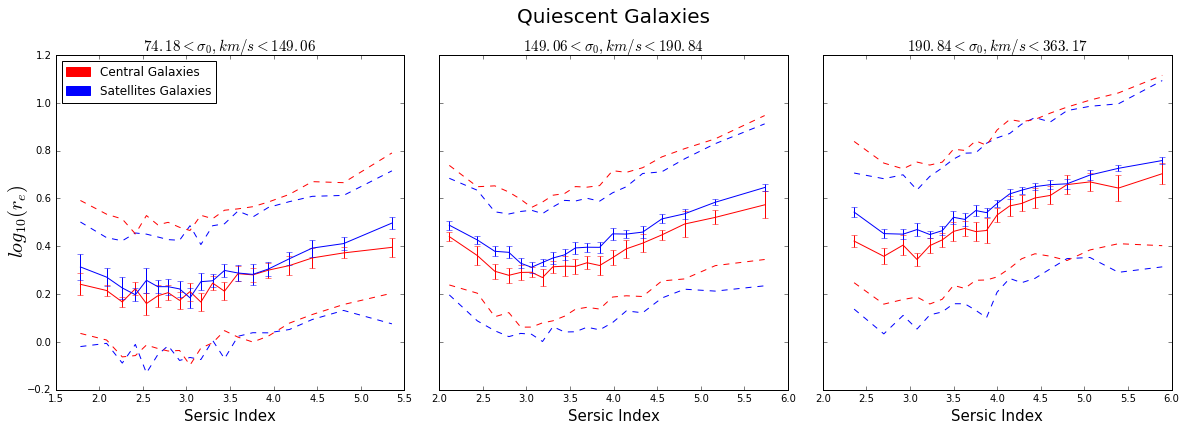}
\centering
\caption{\small
We show the Sersic index-size relations for quiescent central and satellite galaxies in three bins of $\sigma_0$. The error bars are the standard deviation of the medians from 1000 bootstrap resamplings and the dashed lines represent the 1$\sigma$ scatter. These plots show that both central and satellite galaxies are typically larger at higher Sersic Index, which is responsible for the effect of the weights on the size-$\sigma_0$ relation shown in Figure \ref{fig:VoRadSer}.
\label{fig:VoSerRadius}}
\end{figure*}

In this section we investigate whether other correlations between galaxy properties that are independent of environment could be responsible for the observed differences between central and satellite galaxy size and mass, including the star formation rate and morphology distributions.

\subsection{The Effect of Ongoing Star Formation}
\label{OSF}

 Throughout we have split our sample into two separate groups by specific star formation rate, by applying a cut at $log_{10}(sSFR) = -11$. The purpose of this cut was to separate the effect of ongoing star formation within the centrals and satellites, which could lead to a difference in size and mass. As demonstrated by Figures \ref{fig:MassRadius} and \ref{fig:VdispRadius} the star forming galaxies are larger than their quiescent counterparts and so there is a relationship between size and sSFR at fixed mass or $\sigma_0$. As a result, we want to be sure that within the two sets of galaxies there is not a residual difference in the star formation rates between the central and satellite galaxies, which may be responsible for the size and mass\footnote{Star forming galaxies also have higher mass at fixed $\sigma_0$ e.g. \cite{2015ApJ...799..148B}, Figure \ref{fig:VdispMass}} difference between centrals and satellites. For example, if the median specific star formation rate of quiescent galaxies was $10^{-11.5} yr^{-1}$ and the median for satellites was $10^{-12} yr^{-1}$ the central galaxies could be larger simply due to correlation between size and sSFR.

In Figure \ref{fig:SFRVDisp} we compare the distributions of specific star formation rates for central and satellite galaxies, in bins of core velocity dispersion. We also mark on the median sSFR rates in the star forming and quiescent groups. This figure shows that whilst the shapes of the sSFR distributions are pretty similar the central galaxies have a slightly higher median sSFR rate than the satellites. To make sure that these differences are not responsible for any of the previously observed trends we can weight each satellite galaxy in such a way as to make the central and satellite sSFR distributions identical. To do this we weight the satellites in each sSFR bin by the fractional difference between the number of satellite and central galaxies in that bin. We can then add these weights back into our calculations of the differences in size and mass from Section \ref{results}.

We show the weighted results for the size-$\sigma_0$ relation for quiescent galaxies in Figure \ref{fig:VoRadSFR}, which shows that the sSFR weights do not significantly alter the relation or the fractional difference in size, which ranges from 3\% at low dispersion to 29\% at high dispersion. We do not show the star forming galaxies or the mass-$\sigma_0$ relation, but neither of these are significantly affected by the sSFR weights either. These results indicate that despite the central galaxies having slightly higher sSFR than the satellites, even when split into star forming and quiescent subsamples, and there being a correlation between sSFR and size at $\sigma_0$, the differing sSFRs are not contributing to the observed difference in the size or mass of central and satellite galaxies at fixed $\sigma_0$.

\subsection{Differing Morphologies of Centrals and Satellites}

While it is generally true that star forming galaxies are mostly late-type spirals, and quiescent galaxies are predominantly early-types, there may be a difference in the distribution of these morphologies between the centrals and satellites. For instance one could imagine a larger fraction of late-type galaxies amongst quiescent satellites than quiescent centrals as the star formation is halted in satellite galaxies. As morphological late-type galaxies are typically larger at fixed mass and higher mass at fixed $\sigma_0$ than early-type galaxies (both among star forming and quiescent populations e.g. Bezanson et al. 2015) it would follow that, much like in Section \ref{OSF}, any difference in the morphological distribution between central and satellites could be contributing to the observed difference in the size-$\sigma_0$ or mass-$\sigma_0$ relations.

We study this in the same way as we did for the ongoing star formation rates by analysing the S\'ersic indices of the central and satellite galaxies at fixed $\sigma_0$. In Figure \ref{fig:SerVDisp} we show the distributions of morphologies in three bins of $\sigma_0$ for the quiescent galaxies. The satellite galaxies have lower median S\'ersic indices, indicating that they are on average more disk like. Paradoxically, this would actually indicate that the satellites should be larger at fixed mass than the centrals, though the effect may be small enough that it is simply washed out by the environmental signal. The S\'ersic Index is most different for low dispersion galaxies, the difference drops for medium dispersion galaxies and rises again for the high dispersion galaxies.

Much like in the previous section, we wish to test how much the difference in galaxy shape can affect the relations we have found. We build a set of weights for the satellite galaxies by finding the fractional difference between the number of satellites and centrals in each S\'ersic index bin. We apply these weights to the satellite galaxies and recalculate the differences in size and mass at fixed $\sigma_0$. In Figure \ref{fig:VoRadSer} we show the median size at fixed $\sigma_0$ for quiescent centrals and satellites and the difference in size between those galaxies. Comparing to Figure \ref{fig:VdispRadius}, we see that the difference in size has decreased slightly, to $0 \pm 8 \%$ at the lowest $\sigma_0$ to $29 \pm 4 \%$ in the highest $sigma_0$ bin for the random sample. The average difference in size drops to $17\%$ for the mass selected central sample, $16\%$ in the randomly selected central sample and $13\%$ in the $\sigma_0$ selected sample. Interestingly, this is the opposite effect that might be expected. Since satellites are more disk one would have expected them to made them larger at fixed $simga_0$ and thus the difference in size smaller before we corrected for the difference in morphologies. However, as we show in Figure \ref{fig:VoSerRadius} the higher S\'ersic index galaxies are in fact larger and since they receive larger weights from the correction the satellites are made larger, thus the difference in size decreases.

The largest change in the difference in size happens in the lowest dispersion bin which drops from $~5\%$ to almost no difference at all, this is also where the difference in median S\'ersic index is largest in Figure \ref{fig:SerVDisp}. The smallest changes happen at medium dispersions and the change increases again at high dispersion, this mirrors the differences in median S\'ersic Index between central and satellite galaxies.

\section{Summary \& Discussion}
\label{Conclusions}

We have analysed the properties of central and satellite galaxies in a sample of 130,000 galaxies, from Data Release 7 of the SDSS. We use derived quantities from the NYU-VAGC, the MPA/JHU Catalog and the UPenn Photodec Catalog. We split our galaxies into star forming and quiescent groups and then divide them into centrals and satellites using the \cite{2009ApJ...695..900Y} group catalog.

We began by revisiting the size-mass relation of central and satellite galaxies. Using the median half-light radius in bins of fixed stellar mass we observed that for star forming and quiescent galaxies the size-mass relation does not depend on whether a galaxy is a central or a satellite, echoing the results of previous studies \citep[for example,][]{2013ApJ...779...29H}. However we posit that comparing galaxies at fixed mass could be failing to capture the importance of environment in a galaxy's evolution, as mass evolution can also be changed by environmental processes. With mergers affecting central galaxies more often than satellites and satellites being subject to tidal and ram pressure stripping, galaxy harassment and quenching, the net effect will result in central galaxies being more massive than satellites.  Therefore, by comparing galaxies at fixed mass we are not comparing galaxies that were similar prior to a satellite's accretion onto a new dark matter halo and thus we are not capturing the effect of the environments they are being accreted onto. To mitigate these differences, we choose to study the size and mass of galaxies at fixed core velocity dispersion, as events that can change $\sigma_0$ (e.g. major mergers) are rare on the timescales of satellite infall.

We study the size-$\sigma_0$ and mass-$\sigma_0$ relations by finding the median half-light radii and stellar masses within bins of core velocity dispersion. We find that at fixed $\sigma_0$ the central galaxies are consistently larger and more massive than their satellite counterparts by $\sim$15\% on average.  In Section \ref{Systematics} we show that this difference in size and mass is not due to a residual difference in ongoing star formation or morphology between satellites and centrals despite quiescent and star forming satellite galaxies having lower median sSFR and Sersic indices than their central counterparts. 

For star forming galaxies we find that there is very little difference in the size or mass of central and satellite galaxies at fixed $\sigma_0$. Centrals are just a few percent more massive and show almost no size difference except at the lowest dispersions. The lack of much environmental difference is likely the result of star forming satellites having spent a relatively short amount of time as satellites, they are yet to be quenched. A short time as a satellite in a halo means that a galaxy will have continued to form stars at broadly the same rate as an equivalent star forming central galaxy and will be much less likely to have experienced any tidal stripping or disruption. For quiescent galaxies we do see a significant difference between both the size and mass of central and satellite galaxies with central galaxies being both larger and more massive at fixed $\sigma_0$. We also find that there is a strong dependence of the size difference on $\sigma_0$, with low $\sigma_0$ centrals being just a few percent larger and high $\sigma_0$ centrals being up to 30\% larger than equivalent satellites. The mass difference also shows a trend with $\sigma_0$ although it is much weaker.

We have measured the radial mass profiles the central and satellite galaxies at fixed $\sigma_0$. These profiles show that at fixed dispersion, and in all the star formation rate groups, the cores of the central and satellite galaxies are almost identical. However, at larger radii the galaxies begin to differ, and in the outskirts of the galaxies we begin to see larger differences in total mass and stellar mass density. As with the size-$\sigma_0$ and mass-$\sigma_0$ relations, the differences in mass and mass density are greatest for the quiescent galaxies and there is a larger difference for the high dispersion galaxies. These profiles are clear evidence that the satellites have less mass at larger radii than the centrals, while the cores of the galaxies are being preserved against any environmental processes.

Finally we study the mass-metallicity and $\sigma_0$-metallicity relations of central and satellite galaxies. We find that when split by star formation rate there is almost no difference in the  $\sigma_0$-metallicity relations for central and satellite galaxies and at most a 5\% difference in the mass-metallicity relation such that satellites are more metal rich. The small observed difference between central and satellite metallicity at fixed mass is entirely consistent with the difference in mass measured at fixed $\sigma_0$ and the central galaxy mass-metallicity relation. This is consistent with the picture that both the central metallicty and $\sigma_0$ are unaffected by any processes operating differently on central and satellite galaxies whereas the stellar mass is changing.

Under the assumption that $\sigma_0$ is largely conserved when a galaxy becomes a satellite there are a number of ways that the results we've found could occur. Firstly, quiescent central galaxies could continue to accrete mass onto their outskirts whereas their satellite counterparts do not. Alternatively, quiescent satellite galaxies (that were either accreted as quiescent or SF centrals) are being stripped down in size and mass from the outside in. Finally, star forming central galaxies that become quiescent after becoming satellites have different masses and sizes at fixed $\sigma_0$ than quiescent centrals. These processes (or their combination) not only need to produce larger and more massive centrals than satellites but also the variation in those differences with $\sigma_0$.

We expect that all three processes must occur to some degree. Quiescent centrals experience minor mergers which deposit mass preferentially on to the outskirts. As a result of the shape of the stellar-to-halo-mass relation, the more massive (higher $\sigma_0$) the galaxy the more mass they accrete in this way and the larger the fraction of the mass comes from minor mergers. The larger the merger ratio the more the size of a galaxy is expected to increase per unit mass deposited \citep[]{2009ApJ...699L.178N}, and so this naturally explains the increasing size and mass difference between central and satellites as $\sigma_0$ increases. We expect more mass to be accreted onto higher $\sigma_0$ centrals with a larger fraction coming from minor mergers causing a relatively larger size difference. This effect may also be revealed in the size-mass (or size-$\sigma_0$) relation for quiescent galaxies. The slope is steeper at high mass ($\sigma_0$) and shallow at low mass ($\sigma_0$), so for a given change in mass a galaxy grows more at high mass ($\sigma_0$) and less at low mass ($\sigma_0$).

Tidal stripping of stellar mass from satellite galaxies is responsible for the intra-cluster and group light that makes up a significant fraction of stellar mass in groups and clusters. Estimates of the fraction of mass stripped from simulations range from just a few percent up to $\sim$20\% with the amount of mass stripping decreasing with stellar mass and increasing with halo mass \cite[]{2016arXiv160903379B}. The typical fraction of mass lost to stripping is in reasonable agreement with the typical mass difference we observe ($\sim$15\%) and the stripping is expected to occur from the outside in, again consistent with our observed mass profiles. However, we see no dependence of central-satellite mass difference on $\sigma_0$, which appears inconsistent with the expected mass dependent stripping fraction from the models.  


Our final mechanism that could contribute to the difference in the size and mass difference of quiescent central and satellite galaxies is the possibility that there are differing relationships between $\sigma_0$ and stellar mass or size for galaxies that become quiescent while centrals and galaxies that become quiescent while satellites. Quiescent satellites will consist of galaxies that became quiescent as centrals before becoming satellites, and galaxies that became quiescent as a result of becoming satellites. Figures \ref{fig:VdispRadius} and \ref{fig:VdispMass} show that quiescent centrals are $\sim$30\% smaller and less massive than SF centrals with the same $\sigma_0$. If galaxies that become quiescent as satellites evolve more like the SF population of centrals than the quiescent population at fixed $\sigma_0$, continuing to form stars for some time after accretion \cite[e.g.][]{2013MNRAS.432..336W,2016MNRAS.463.3083O}, they may be more like their SF central counter parts than their quiescent central ones, i.e. more massive and larger. This would tend to counter act any effect of tidal stripping of satellites or central growth from minor mergers. Since the fraction of satellite galaxies that were SF when they became satellites will increase as $\sigma_0$ decreases, a result of the strong correlation between quiescence and $\sigma_0$ \cite[e.g.][]{2012ApJ...751L..44W,2016MNRAS.462.2559B}, we would expect any difference in size and mass at fixed $\sigma_0$ between satellite galaxies that became quiescent as satellites, rather than as centrals, to mainly affect the lower $\sigma_0$ population. We anticipate that the corrections we have applied for the small differences in SFR and S\'ersic index between the central and satellite quiescent galaxies in Section \ref{Systematics} would mitigate these effects, but it remains possible that the differing quenching mechanisms and/or timescales could mean that satellites quenched as satellites could still be larger and more massive than satellites quenched as centrals, even when their SFRs and S\'ersic indices are matched. 

Taken together, these considerations seem to point to the continued growth of central galaxies by minor mergers being largely responsible for the mass and size differences between quiescent central and satellite galaxies at fixed $\sigma_0$, particularly at high $\sigma_0$, with tidal stripping playing a minor role that increases as $\sigma_0$ decreases. Since at high $\sigma_0$ almost all galaxies are quiescent \citep[][]{2012ApJ...751L..44W,2016MNRAS.462.2559B}, and tidal stripping is expected to be minimal, it is reasonable to assume that essentially all the difference in size (25\%) and mass (20\%) of central and satellite galaxies is caused by minor mergers on to the centrals that have occurred over the average time that the satellites have been satellites \citep[$\sim$3 Gyr][]{2013MNRAS.432..336W}.

\section*{Acknowledgements}

We would like to thank Rachel Bezanson, Marijn Franx and Andrew Wetzel for enlightening discussions and comments on early versions of this work, Britt Lundgren for a careful reading of the final manuscript, and the anonymous referee for their help comments and suggestions. We would also like to thank all of the authors who compiled and publicly released the value added catalogs (NYU VAGC, MPA/JHU catalogs, UPenn SDSS PhotDec catalog, Yang et al. Group catalog) we used in this work. Without those efforts, and those of the SDSS generally, to make their data publicly available in easily used forms this work would be close to being impossible.
 
Funding for the SDSS and SDSS-II has been provided by the Alfred P. Sloan Foundation, the Participating Institutions, the National Science Foundation, the U.S. Department of Energy, the National Aeronautics and Space Administration, the Japanese Monbukagakusho, the Max Planck Society, and the Higher Education Funding Council for England. The SDSS Web Site is http://www.sdss.org/.

The SDSS is managed by the Astrophysical Research Consortium for the Participating Institutions. The Participating Institutions are the American Museum of Natural History, Astrophysical Institute Potsdam, University of Basel, University of Cambridge, Case Western Reserve University, University of Chicago, Drexel University, Fermilab, the Institute for Advanced Study, the Japan Participation Group, Johns Hopkins University, the Joint Institute for Nuclear Astrophysics, the Kavli Institute for Particle Astrophysics and Cosmology, the Korean Scientist Group, the Chinese Academy of Sciences (LAMOST), Los Alamos National Laboratory, the Max-Planck-Institute for Astronomy (MPIA), the Max-Planck-Institute for Astrophysics (MPA), New Mexico State University, Ohio State University, University of Pittsburgh, University of Portsmouth, Princeton University, the United States Naval Observatory, and the University of Washington.




\bibliographystyle{mnras}
\bibliography{VDisp} 







\bsp	
\label{lastpage}
\end{document}